\documentclass[cits]{PoS}

\usepackage{amsmath,amssymb}

\providecommand{\abs}[1]{\left\lvert#1\right\rvert}

\newcommand{\expv}[1]{\left\langle#1\right\rangle}

\title{\vspace*{-5.5cm}
{\hfill \texttt{\footnotesize CERN-PH-TH-2015-266}}
\vfill
Effects of higher dimension
operators on the Standard Model Higgs sector}
\ShortTitle{Effects of higher dimension
operators on the SM Higgs sector}

\author{\speaker{Oscar \AA{}kerlund}\\
        Institut f\"ur Theoretische Physik, ETH Zurich, CH-8093 Z\"urich, Switzerland\\
        E-mail: \email{oscara@itp.phys.ethz.ch}}

\author{Philippe~de~Forcrand\\
        Institut f\"ur Theoretische Physik, ETH Zurich, CH-8093 Z\"urich, Switzerland\\
        CERN, Physics Department, TH Unit, CH-1211 Geneva 23, Switzerland\\
        E-mail: \email{forcrand@itp.phys.ethz.ch}}

\author{Jakob Steinbauer\\
        Institut f\"ur Theoretische Physik, ETH Zurich, CH-8093 Z\"urich, Switzerland\\
        E-mail: \email{jakobste@student.ethz.ch}}

\abstract{We study the effect of higher dimension operators on the electroweak finite
  temperature phase transition in two sectors of the Standard Model. Firstly, the
  Higgs-Yukawa sector, consisting of the Higgs doublet and the massive Standard Model fermions,
  is studied with an approximate method, Extended Mean Field Theory. Secondly,
  the gauge-Higgs sector, consisting of the Higgs doublet and the gauge fields of the
  weak interaction, is studied using Monte Carlo simulations. In both cases we find that a
  cutoff scale of around 1.5~TeV is needed to make the electroweak phase transition first
  order at the experimental value of the Higgs boson mass, which is a requirement for making
  electroweak baryogenesis viable.}

\FullConference{The 33rd International Symposium on Lattice Field Theory\\
		14 -18 July 2015\\
		Kobe International Conference Center, Kobe, Japan}

\graphicspath{{figures/}} 
\begin{document}

\section{Introduction}\noindent
Despite its outstanding success, the Standard Model (SM) is incomplete\footnote{Most obviously due to the
  fact that it is completely neglecting gravity.} and can thus only
be considered an effective model. Due to its renormalizability, and the (seemingly)
special value of the Higgs boson mass, however, its range of validity seems to be
huge and it may even be internally consistent all the way up to the Planck scale~\cite{Espinosa,Branchina}.
This is both intriguing and challenging, since any UV completion of the SM must leave most
of its observables almost unchanged at the same time as it solves the problems of the SM,
such as dark matter, neutrino oscillations and baryogenesis. Yet, although the SM does not
require new physics to be important until very high energies, there is no general reason
to exclude new degrees of freedom with masses around a few TeV. Something similar happens
in the Fermi Theory of Weak interactions where the mass of the $W$-boson of $80$~GeV is
significantly below $4\pi{}\expv{\phi}\approx 3$~TeV, where the theory stops being perturbative.

In this proceeding we study generic effects of Beyond the Standard Model (BSM) physics
at intermediate scales by including the dimension six operator $\abs{\phi}^6$ in the SM Higgs
sector. Generally, there are of course many more higher order operators~\cite{Grzadkowski:2010es} but for simplicity we will let
$\abs{\phi}^6$ serve as a proxy for all of them. The main focus will be on the order of the electroweak
finite temperature transition as a function of the Higgs mass $M_h$ and of the coefficient of the dimension six operator,
parametrically given by $M_{\rm{BSM}}^{-2}$, where $M_{\rm BSM}$ can be thought of as the mass of the lightest particle in the
UV completion of the SM which couples to the Higgs boson. The motivation behind this is to investigate the possibility of
electroweak baryogenesis, which, due to one of the three Sakharov conditions for baryogenesis,
can only be viable if there is a first order transition in the electroweak sector as the universe
cools down. With no dimension six operator, the transition is a crossover for the Higgs mass
at its experimental value $125$~GeV, and it would only become first order for a Higgs mass below about
$70$~GeV~\cite{Rummukainen:1998as}. However, in the presence of a $\abs{\phi}^6$ operator the critical value of the Higgs mass
can be raised and it is interesting to determine for which value of $M_{\rm{BSM}}$ the critical
Higgs mass exceeds the experimental value. Since it is not known how to represent the full SM on the
lattice we answer this question in two different simplified versions of it: first in the Higgs-Yukawa
model which neglects all gauge fields but retains the Higgs field and all (massive) SM fermions, and then
in a gauge-Higgs model consisting of the Higgs field and the $SU(2)$ gauge fields.

\section{Higgs-Yukawa model}\noindent
The Higgs-Yukawa model neglects the gauge degrees of freedom of the SM and consists of the complex, two component,
scalar Higgs field and the massive SM fermions, coupled to the Higgs field via Yukawa interaction terms. Since,
at tree level, the Yukawa couplings $y_f=M_f/\expv{\phi}$ are proportional to the masses of the fermions $M_f$,
and since the top quark mass $M_t$ is of the same order of magnitude as the Higgs expectation value $\expv{\phi}$,
it is important to treat the model nonperturbatively. Due to the chiral nature of the Yukawa coupling it is also
important to use a chiral Dirac
operator on the lattice, for example the Neuberger Overlap operator. This in turn makes full, dynamical lattice
simulations of the model very demanding and expensive in terms of computer time. We will follow a different route and
solve the model approximately using Extended Mean Field Theory (EMFT) which comprises a self-consistent determination
of the Higgs expectation value and self-energy due to the self- and Yukawa-interactions. In this approximation the fermions
couple to the magnitude $\abs{\varphi_0}$ of the Higgs field, which is taken to be constant in space time
so that the Dirac operator $D^{\rm ov}$ factorizes into independent blocks, one for each fermion. The magnitude varies,
however, as we integrate over all the components of the Higgs field. The lattice action takes the form
\begin{align}
  S &= \sum_x\Bigg\{-\kappa\sum_\mu\varphi^\dagger_x\varphi_{x+\hat\mu}+\text{h.c.} + \abs{\varphi_x}^2 +
  4\kappa^2\lambda\left(\abs{\varphi_x}^2-1\right)^2 + 8\kappa^3\lambda_6\abs{\varphi}^6 \nonumber\\
  &\phantom{=\sum_x\Bigg\{}+\sum_fN_{c,f}\text{TrLog}\left(D^{\rm ov}(y_f\sqrt{2\kappa}\abs{\varphi_0})\right)\Bigg\},\label{eq:action_trlog}
\end{align}
where $\kappa$ is inversely proportional to the bare mass squared and where $\lambda_6 = (aM_{\rm BSM})^{-2}$.
$N_{c,f}$ is a multiplicity factor which equals $3$ for quarks (number of colors) and $1$ for leptons.

\begin{figure}[htp]
\centering
{\scriptsize\hspace{5mm}$\lambda_6=0.1$\hspace{7cm}$\lambda_6=1$}\\[-2mm]
\includegraphics[width=0.49\linewidth]{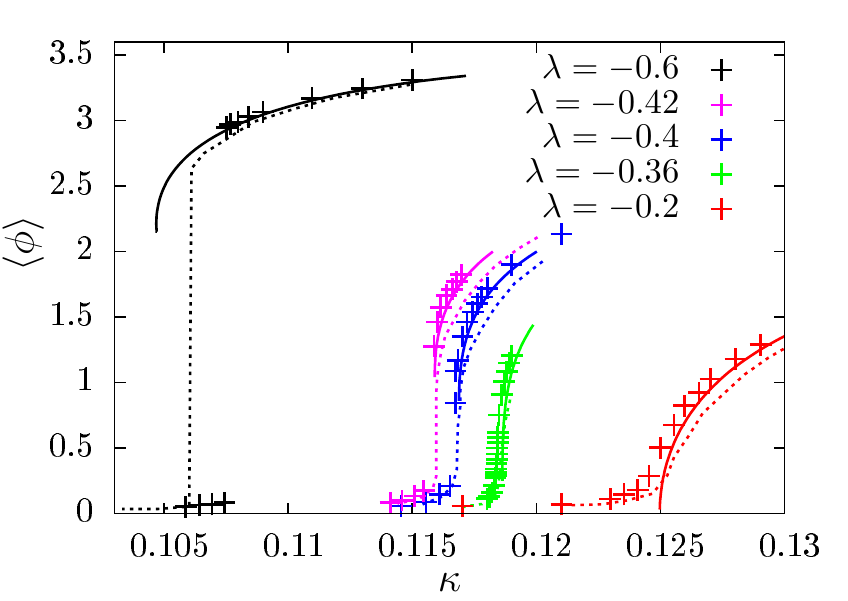}
\includegraphics[width=0.49\linewidth]{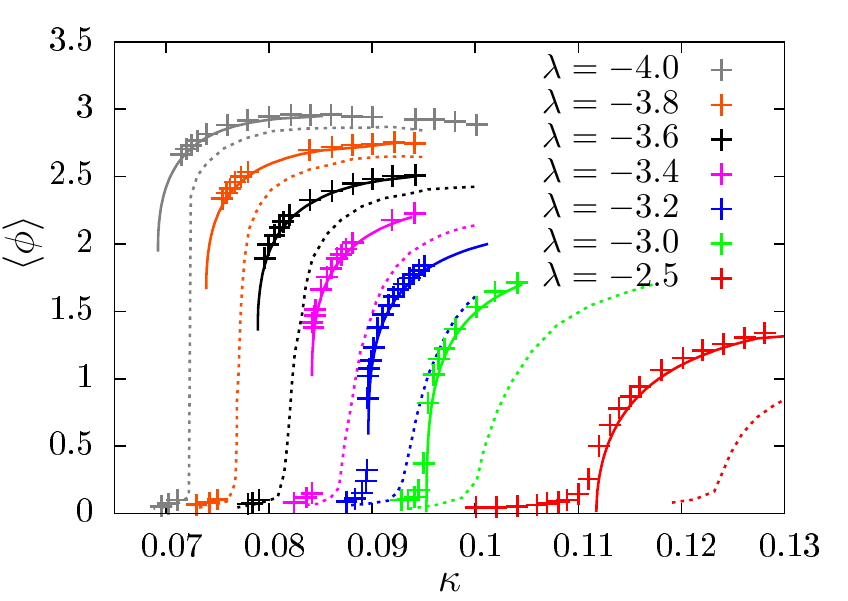}
\caption{The Higgs expectation value $\expv{\phi}$ in lattice units for various quartic couplings $\lambda$ with $\lambda_6=0.1$
  (\emph{left panel}) and $\lambda_6=1$ (\emph{right panel}). The symbols and dashed lines are Monte Carlo data and
  Constraint Effective Potential data respectively, taken from~\cite{Chu:2015nha,Jansen}. The solid lines are the EMFT results
  obtained in this work.}
  \label{fig:comparison}
\end{figure}

This action is then solved using EMFT, see \cite{Akerlund:2015fya} for details, and in particular we determine the
finite temperature phase diagram in the $(M_{\rm BSM},M_h)$-plane. As a consistency check, we first compare the EMFT
solution to full Monte Carlo simulations~\cite{Chu:2015nha,Jansen}, which have been performed using a restricted version of the model
with only the top and bottom quarks, which were further taken to be mass degenerate (otherwise there is a sign problem),
and unit multiplicity factors $N_{c,f}$. In Fig.~\ref{fig:comparison} we show the Higgs expectation value $\expv{\phi}$ as a function
of the hopping parameter $\kappa$ for various values of the quartic coupling $\lambda$ at a small value of $\lambda_6=0.1$
(\emph{left panel}) and at a larger, nonperturbative value of $\lambda_6=1$ (\emph{right panel}). The symbols are
Monte Carlo data and the dashed lines are obtained by a perturbative effective model called the Constraint Effective
Potential, both taken from~\cite{Chu:2015nha,Jansen}. The solid lines are the EMFT results from this work. Clearly EMFT is an
excellent approximation at all parameter values, in addition to being orders of magnitude cheaper computationally than
the Monte Carlo simulations.

\begin{figure}[htp]
\centering
\includegraphics[width=0.49\linewidth]{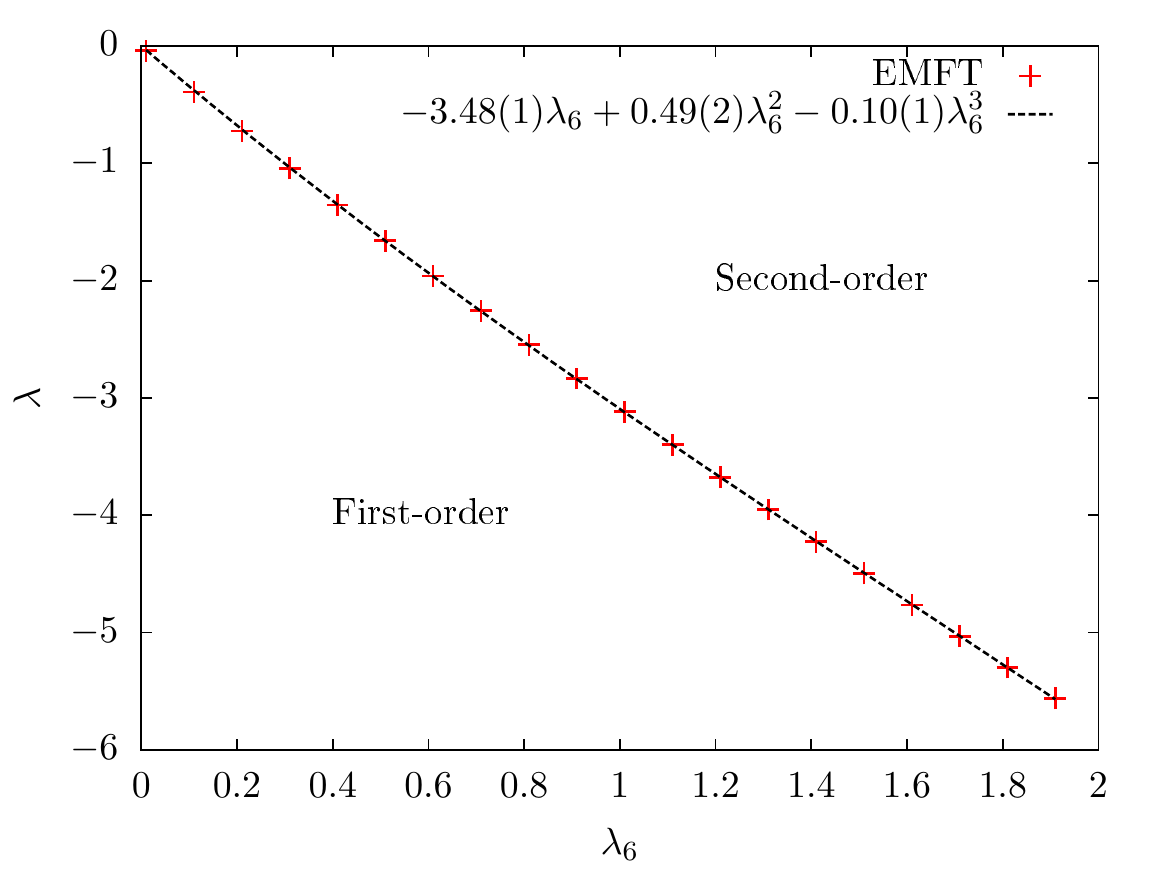}
\includegraphics[width=0.49\linewidth]{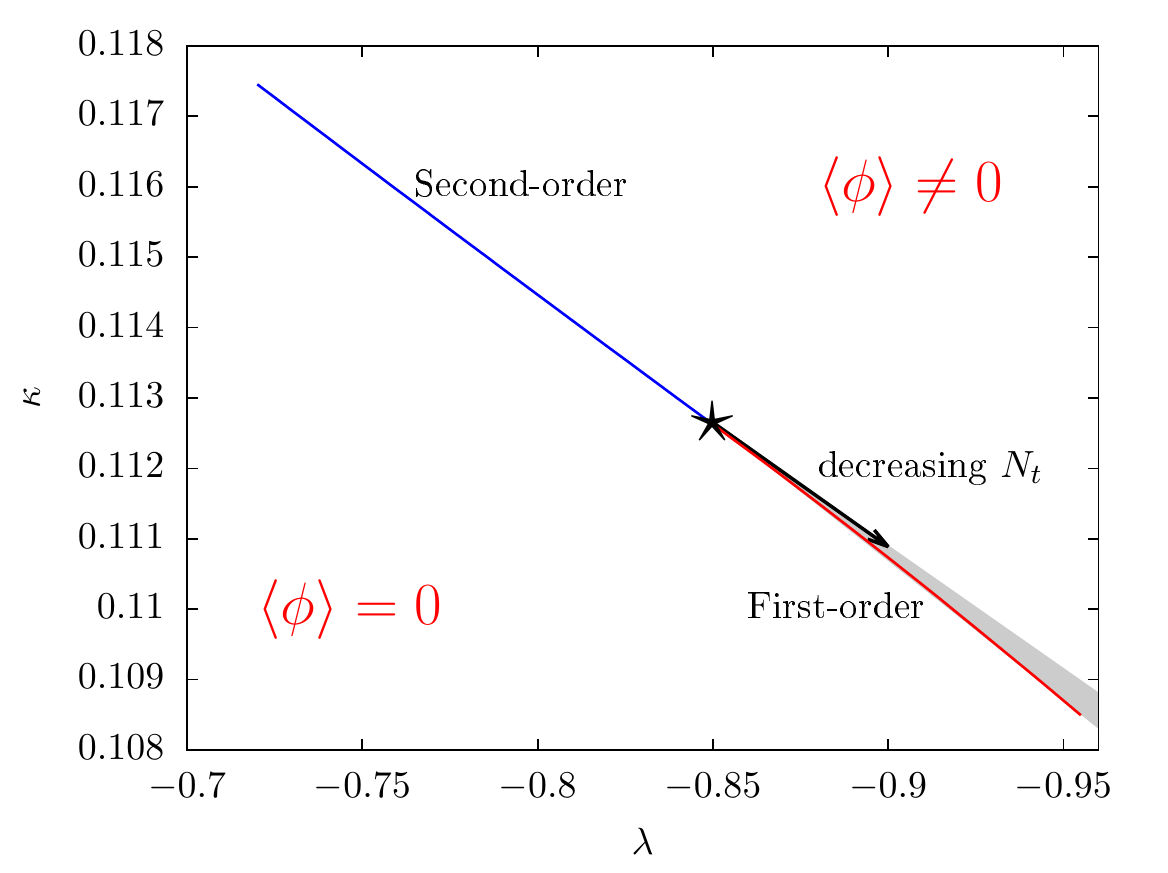}
\caption{\emph{Left panel}: The tricritical line at zero temperature in the $(\lambda_6,\lambda)$-plane.
  For $\lambda$ below the line the transition is first order and above it is second order. \emph{Right panel}:
  The phase diagram in the $(\lambda,\kappa)$-plane for $\lambda_6=1/4$ at zero temperature. On the blue line the
  transition is second order and on the red line it is first order. The star in between is the tricritical point. As the
  temperature is increased the tricritical point moves along the black arrow and the first order line sweeps out the
  gray area which is where a first order finite temperature transition can be found.}
  \label{fig:first_order_line}
\end{figure}

In order to obtain the full phase diagram in physical parameters, we first determine the order of the $\kappa$-driven
transition in the $(\lambda,\lambda_6)$-plane at zero temperature. Without gauge fields the expectation value of the
Higgs field is a true order parameter so the transition can be either of first or second order as seen in
Fig.~\ref{fig:comparison}. For any positive value of $\lambda_6$, there is a tricritical value of $\lambda<0$ for
which the transition turns from second to first order as $\lambda$ is taken to be more negative. How this tricritical
coupling depends on $\lambda_6$ can be seen in the left panel of Fig.~\ref{fig:first_order_line}. In the right panel
we have fixed $\lambda_6=1/4$ and we show
the phase diagram in the $(\lambda,\kappa)$-plane. The star denotes the tricritical point and corresponds to the
intersection of the line in the right panel and the line in the left panel of Fig.~\ref{fig:first_order_line}.
We now turn on temperature by making the lattice extent finite in the temporal direction and study the effect on the
tricritical point. Its trajectory as the temperature is increased is indicated by the arrow in the right panel of
Fig.~\ref{fig:first_order_line} and since it moves into the region of the symmetry broken phase, we conclude that the
symmetry restored phase is expanding, as we know it should since at very high temperature the model is always in
the symmetric phase. The interesting region in this phase diagram is the shaded gray area between the red first-order
line and the trajectory of the tricritical point. This area is swept out by the first order line as the temperature
rises and thus, if the
parameters are such that at zero temperature we are in the broken phase somewhere in the gray area, for some finite
temperature the first-order line will traverse this point and we will go to the symmetric phase via a first-order
phase transition. This is exactly the scenario we are looking for, except in reverse as the universe cools down.

\begin{figure}[htp]
\centering
\includegraphics[width=0.49\linewidth]{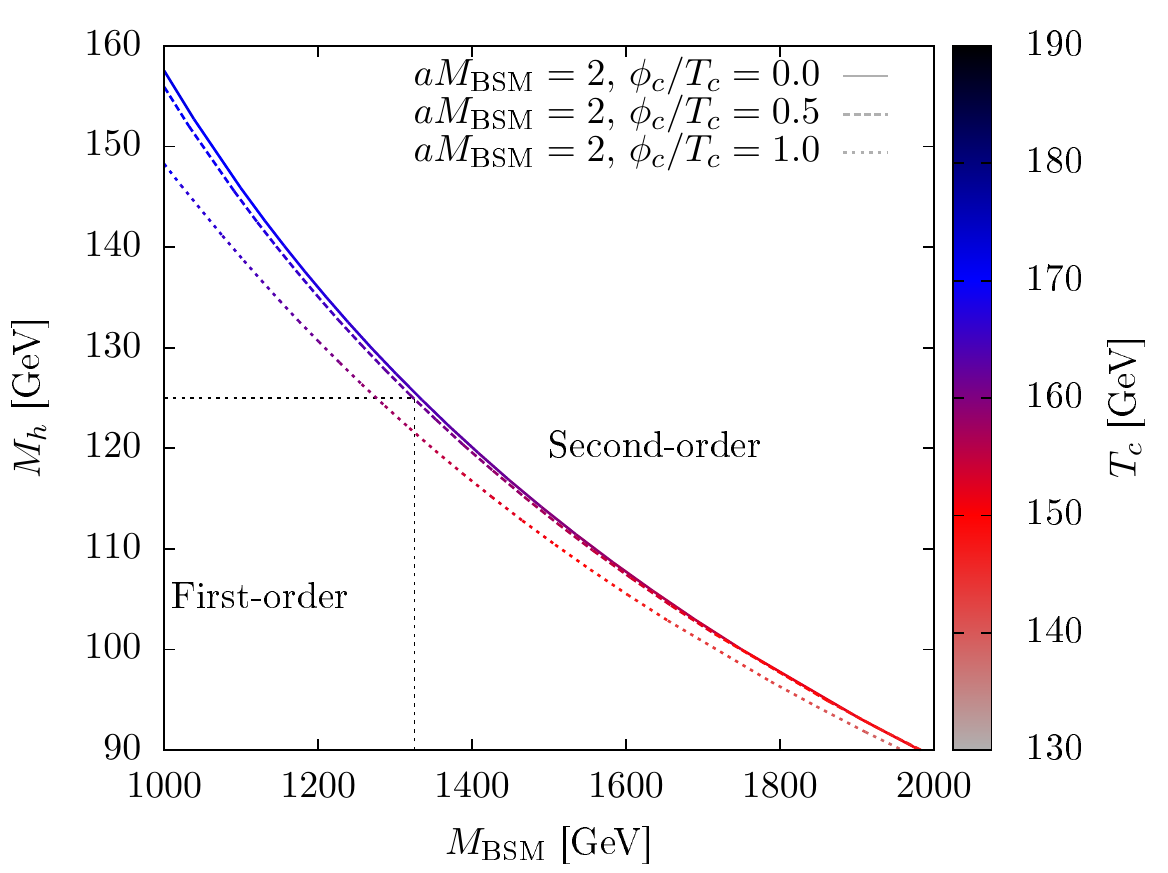}
\includegraphics[width=0.49\linewidth]{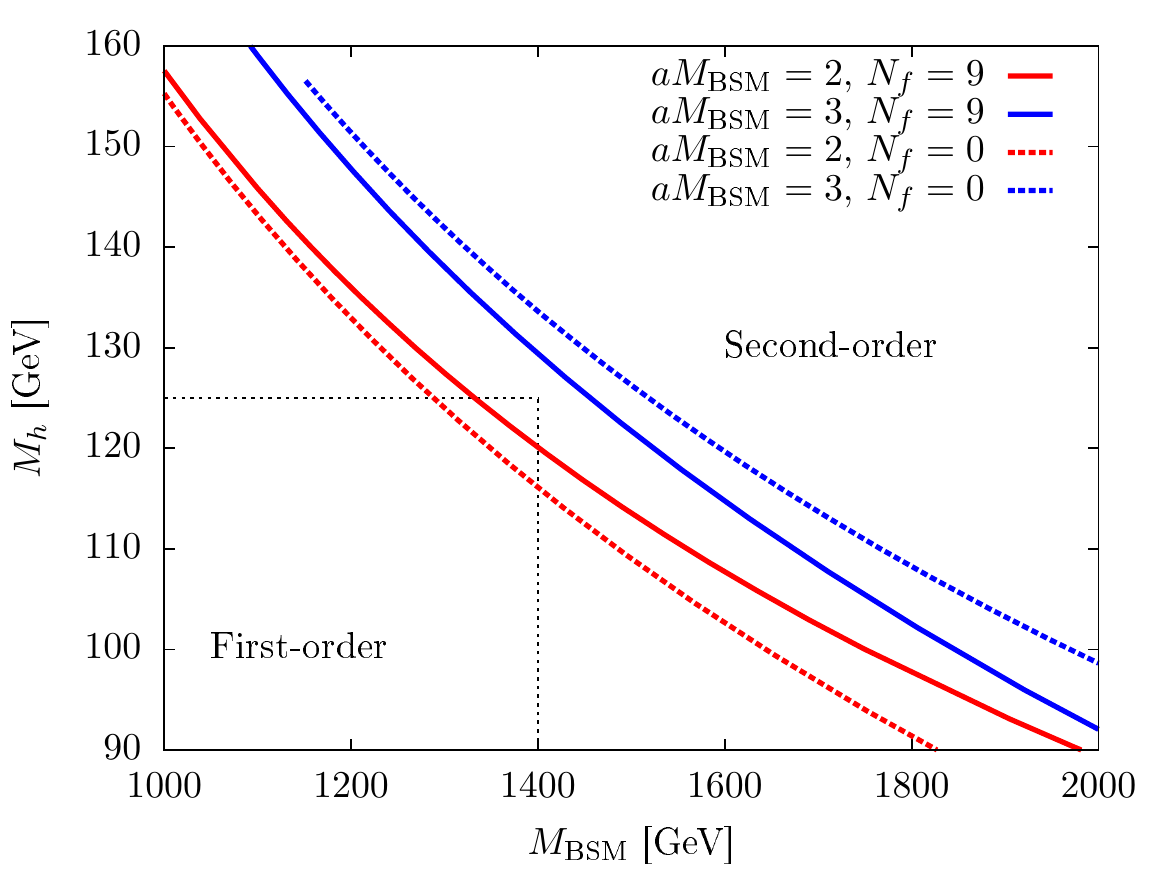}
\caption{\emph{Left panel}: The phase diagram in the $(M_{\rm BSM},M_h)$-plane. The lines are lines of constant $\phi_c/T_c$
  where $\phi_c$ is the expectation value at $T_c$. The solid line where $\phi_c/T_c=0$ is the tricritical line.
  As $M_{\rm BSM}$ is decreased the transition becomes stronger. The color coding gives $T_c$ in GeV.
  \emph{Right panel}: The tricritical line with and without all massive SM fermions at two different values of $aM_{\rm BSM}$.
  The spread of these lines gives an estimate of the systematic uncertainties of the effective model. The effect of fermions
  is rather small.}
  \label{fig:phase_diag_HY}
\end{figure}

The last step consists of considering curves of fixed $\phi_c/T_c$ inside the gray area, i.e. curves where the expectation
value of the field $\phi_c$ at the critical temperature $T_c$ is constant in units of the critical temperature.
A value of $0$ corresponds to the trajectory of the tricritical point itself and a value of $1$ is generally required
in order to have
a strong enough first-order transition for electroweak baryogenesis to be viable. In the left panel of
Fig.~\ref{fig:phase_diag_HY} we show this curve in the $(M_{\rm BSM},M_h)$-plane for three different values of $\phi_c/T_c$.
The color coding denotes $T_c$ in GeV and the dashed lines show the experimental value of the Higgs mass $M_h$ and a
nominal value for $M_{\rm BSM}$. We have fixed the lattice spacing such that $aM_{\rm BSM}=2$. In the right panel of
Fig.~\ref{fig:phase_diag_HY} we show the impact the fermions have on the result by comparing the trajectory of the
tricritical point using all nine massive SM fermions and that with only the pure Higgs sector. We do this for two different values
of the lattice spacing in terms of $M_{\rm BSM}$ and find that the effect of fermions is rather small and of indefinite sign, indicating
that the physics is dominantly determined by the Higgs self-interaction.

\section{Gauge-Higgs model}\noindent
The gauge-Higgs model consists of the Higgs field and the weak gauge degrees of freedom and thus neglects all fermionic
fields. In the light of the result above this seems well justified. Moreover, it is an interesting model on its own,
where the gauge fields yield important contributions to the couplings in the thermal effective theory of the
Higgs sector~\cite{Csikor:1998eu}. The lattice action for this model is given by
\begin{equation}
  \mathcal{L}_{\mathrm{GH}}=-\sum_\mu\kappa_\mu\phi^\dagger(x) U_\mu(x)\phi(x+\hat\mu) +\abs{\phi(x)}^2
  + \lambda\left(\abs{\phi(x)}^2-1\right)^2 - \frac{1}{2}\sum_{\mu>\nu}\beta_{\mu\nu}\mathrm{ReTr}P_{\mu\nu}(x),
\end{equation}
where $\kappa_\mu$ and $\beta_{\mu\nu}$ are the anisotropic hopping parameter and gauge coupling respectively. These will take
different values for the spatial, $\mu\in\{1,2,3\}$, and temporal, $\mu=4$, directions and we define the bare
anisotropy factors to be $\gamma_\kappa^2 = \kappa_4/\kappa_i$ and $\gamma_\beta^2=\beta_{4i}/\beta_{ij}$. The reason for choosing
anisotropic couplings is that this will allow us to reach higher temperatures without reducing the spatial lattice size
too much and thus the total number of lattice sites can be kept under control. A more complete description of the
model and technical details of the simulations can be found in~\cite{Steinbauer:mt}~and~\cite{Steinbauer:2015}.

At tree level the bare anisotropy factors should be equal since they both represent the ratio $\gamma_\kappa=\gamma_\beta
=a_s/a_t$ of lattice spacings in the spatial and temporal directions. However, due to quantum effects they get
renormalized and acquire different corrections. So in order to have a renormalized anisotropy $\xi=a_s/a_t$
different from one, $\gamma_\kappa$ and $\gamma_\beta$ need to be tuned individually. In the Higgs channel we use the
ratio of spatial versus temporal Higgs masses to tune $\gamma_\kappa$ and in the gauge channel we use ratios of elongated Wilson
loops in spatial versus temporal planes to tune $\gamma_\beta$. For the parameter values
we considered they both obtain small corrections, not too different from 1-loop perturbative values, which however depend
rather strongly on the exact values of the bare couplings.

\begin{figure}[htp]
\centering
\includegraphics[width=0.50\linewidth]{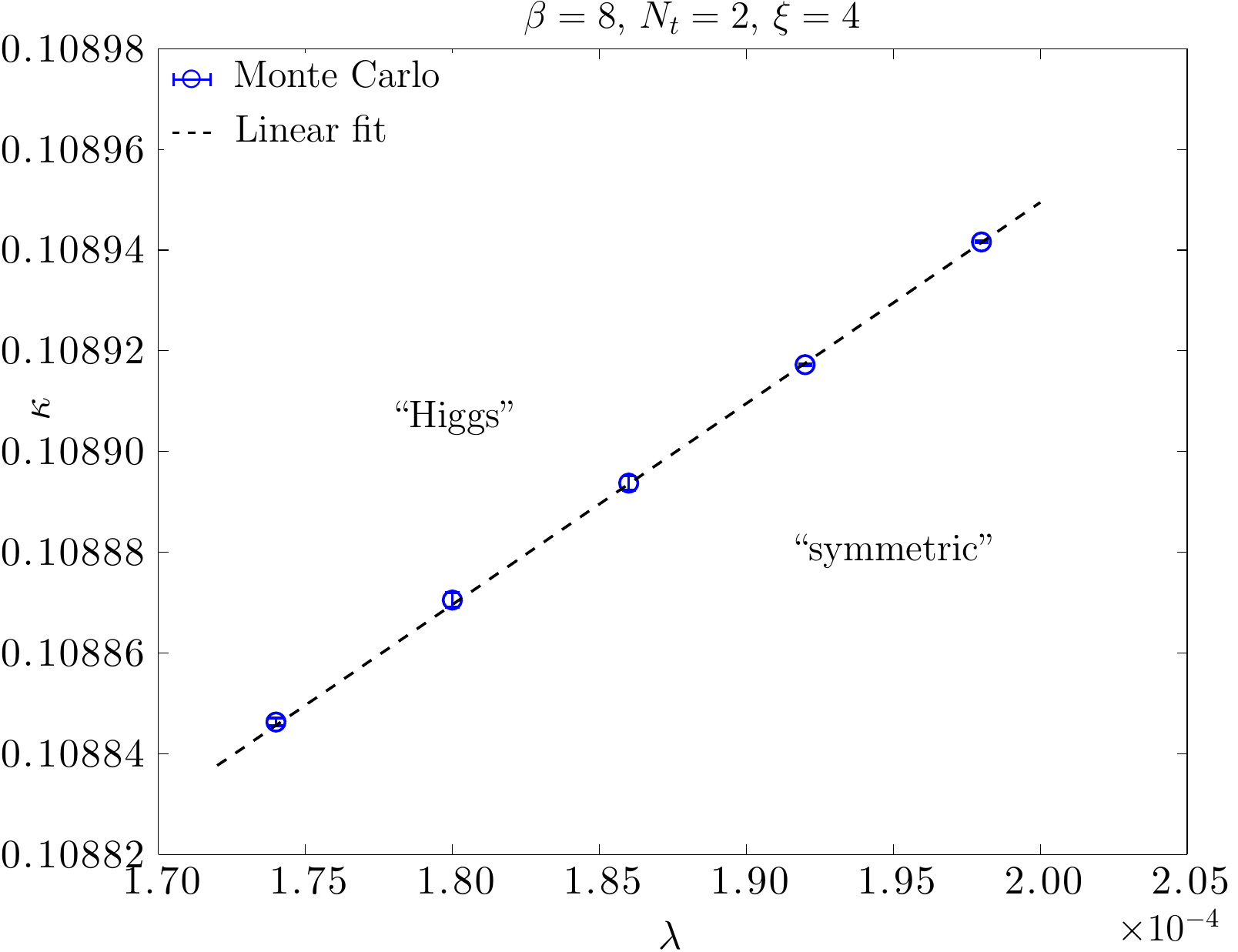}
\includegraphics[width=0.48\linewidth]{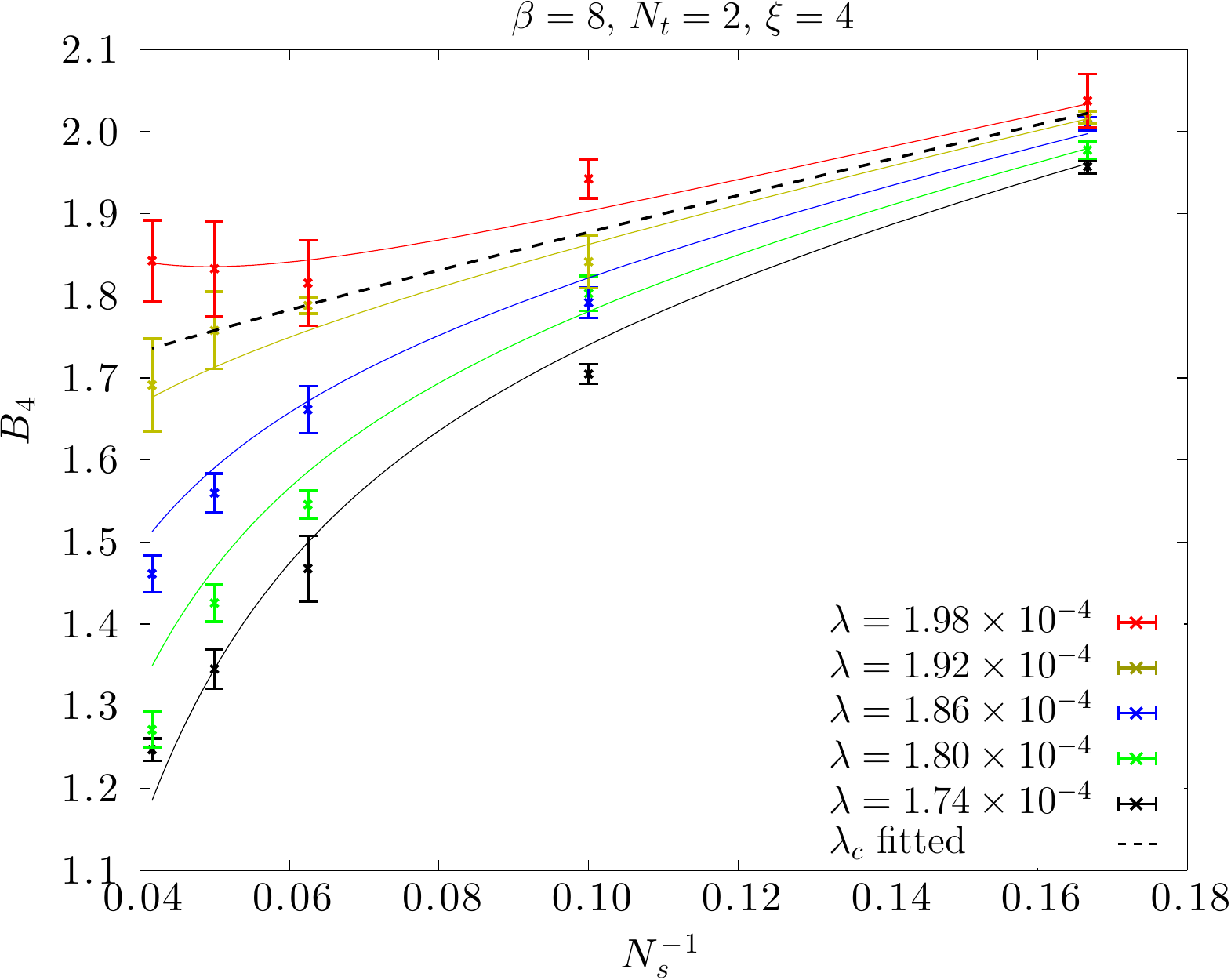}
\caption{\emph{Left panel}: The phase diagram at $\lambda_6=0$. For small values of $\lambda$ the transition is first
  order and for large values it is crossover. \emph{Right panel}: The Binder cumulant $B_4$ at the pseudo-critical point as
  a function of the spatial box size $L$ for different values of $\lambda$. We fit this data to find $\lambda_c$
  where $B_4$ takes the value of the $3d$ Ising model universality class in the thermodynamic limit.}
\label{fig:phase_diag_lam60}
\end{figure}

After the anisotropies have been tuned at zero temperature, we can determine the phase diagram in the
$(\lambda,\kappa)$-plane at finite temperature, which is shown in the left panel of Fig.~\ref{fig:phase_diag_lam60} where
a lattice of size $20^3\times2$ has been used. It is known that
the Higgs mass is increasing with $\lambda$, so we expect that for small $\lambda$ the transition will be first order
and for large $\lambda$ it will be a crossover. Somewhere in between there is a critical value where the transition
is second order, in the universality class of the $3d$ Ising model~\cite{Rummukainen:1998as}.
In the right panel of Fig.~\ref{fig:phase_diag_lam60} we show the spatial size dependence of
the Binder cumulant $B_4$ of the magnetization-like observable, which is used to fit the value of $\lambda$ at the critical point.
The value of the Higgs mass $M_h$ at this point, in units of the $W$ mass, marks the critical value for which a second order transition can
be obtained. In the absence of a $\abs{\phi}^6$ term in the potential, this mass is found to be $67.5(5)$~GeV, in good agreement
with~\cite{Csikor:1998eu}.

\begin{figure}[htp]
\centering
\includegraphics[width=0.48\linewidth]{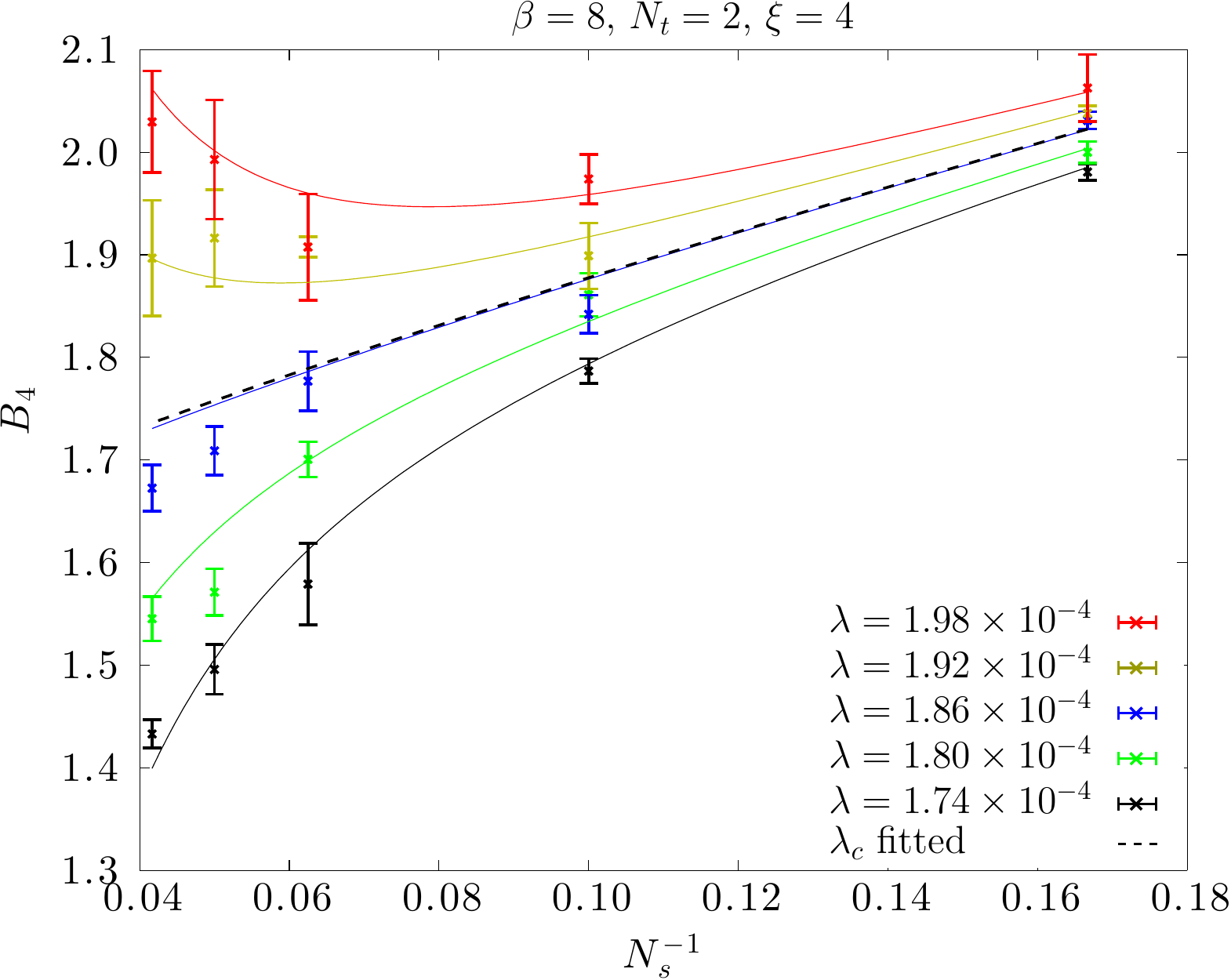}
\includegraphics[width=0.51\linewidth]{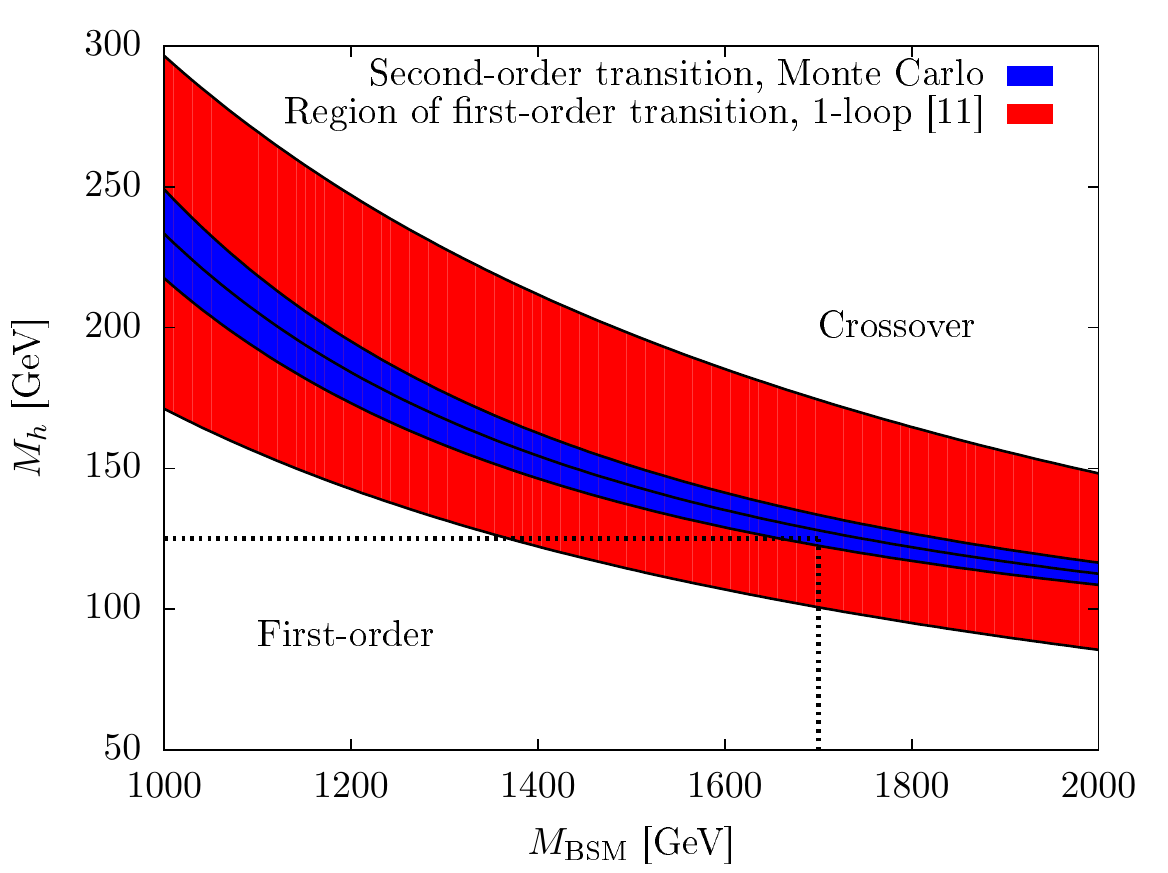}
\caption{\emph{Left panel}: The same as the right panel of Fig.~\protect\ref{fig:phase_diag_lam60} but for $\lambda_6=10^{-7}$.
  \emph{Right panel}: The tricritical line in the $(M_{\rm BSM},M_h)$-plane obtained by extrapolating the linear response
  from small $\lambda_6=M_{\rm BSM}^{-2}$. Above the central line in the blue region the transition is a crossover and below
  it is first order. The red region is obtained from a tree-level evaluation of the Higgs potential with an additional
  one-loop thermal mass~\cite{Grojean}.}
\label{fig:phase_diag_lam6n0}
\end{figure}

We now turn on the $\abs{\phi}^6$ operator by reweighing the ensembles generated without it. Since the location of
the critical point depends on $\lambda_6\propto{}M_{\rm BSM}^{-2}$, the coefficient of $\abs{\phi}^6$, a simultaneous reweighing
in $\kappa$, $\lambda$ and $\lambda_6$ is needed. In the left panel of Fig.~\ref{fig:phase_diag_lam6n0} we show $B_4$
in the case where $\lambda_6=10^{-7}$. The critical $\lambda$ decreases with $\lambda_6$ but due to the positive curvature
generated by the $\abs{\phi}^6$ term the Higgs mass increases with $\lambda_6$. From the linear response of $M_h$ to 
$\lambda_6$ we can determine the trajectory of the tricritical point in the $(M_h,M_{\rm BSM})$-plane, which is shown in
the right panel of Fig.~\ref{fig:phase_diag_lam6n0}. This should be compared with the left panel in
Fig.~\ref{fig:phase_diag_HY}: we see that the two models yield consistent results and that they are completely consistent
with perturbation theory~\cite{Grojean}.

\section{Conclusions}
We have investigated two simplified versions of the Standard Model, the Higgs-Yukawa model and a gauge-Higgs model,
in the presence of a higher dimension operator, $\abs{\phi}^6$, parametrically suppressed by two powers of a ``new
physics'' energy scale, labeled $M_{\rm BSM}$. We have determined the curve in the $(M_h,M_{\rm BSM})$-plane ($M_h$ being
the Higgs mass) where the electroweak finite temperature transition turns first order. This is relevant
in the context of electroweak baryogenesis where a strong first order transition is needed in order to fulfill
all Sakharov conditions. For the experimental value of the Higgs mass, $M_h=125$~GeV, both models require $M_{\rm BSM}$
to be around $1.5$~TeV, which is well within what will be probed by run II of LHC. The fact that both models yield similar
results indicates that the Higgs potential itself is
most important when determining $M_{\rm BSM}$, whereas the gauge and fermion degrees of freedom yield only small corrections,
captured well by perturbation theory.

\end{document}